\documentclass[12pt,preprint]{aastex}
\begin{document}
\newcommand{\msun}{M_{\odot}}
\newcommand{\kms}{\, {\rm km\, s}^{-1}}
\newcommand{\cm}{\, {\rm cm}}
\newcommand{\gm}{\, {\rm g}}
\newcommand{\erg}{\, {\rm erg}}
\newcommand{\kpc}{\, {\rm kpc}}
\newcommand{\mpc}{\, {\rm Mpc}}
\newcommand{\seg}{\, {\rm s}}
\newcommand{\kev}{\, {\rm keV}}
\newcommand{\hz}{\, {\rm Hz}}
\newcommand{\etal}{et al.\ }
\newcommand{\yr}{\, {\rm yr}}
\newcommand{\eq}{eq.\ }
\def\arcsec{''\hskip-3pt .}

\title{Orbital perturbations of transiting planets:
A possible method to measure stellar quadrupoles and to detect Earth-mass
planets}
\author{Jordi Miralda-Escud\'e$^{1}$}
\affil{Ohio State University, Columbus OH 43202}
\affil{Institute for Advanced Study, Princeton NJ 08540}
\email{jordi@astronomy.ohio-state.edu}
\affil{$^{1}$ Alfred P. Sloan Fellow}

\begin{abstract}

  The recent discovery of a planetary transit in the star HD 209458,
and the subsequent highly precise observation of the transit lightcurve
with HST, is encouraging to search for any phenomena that might induce
small changes in the lightcurve. Here we consider the effect of the
quadrupole moment of the parent star and of a possible second planet
perturbing the orbit of the transiting planet. Both of these cause a
precession of the orbital plane and of the periastron of the planet,
which result in a long term variation of the duration and the period of
the transits. For a transiting planet at $0.05$ AU,
either a quadrupole moment similar to that of
the Sun or the gravitational tug from an Earth-like planet on an orbit
of semimajor axis $\sim 0.2$ AU and a relative inclination near the
optimal $45^\circ$ would cause a
transit duration time derivative of $\sim 1$ second per year.

\end{abstract}


\section{Introduction}

  The first transiting extrasolar planet has been identified around
the G0 star HD209458 (Charbonneau \etal 2000; Henry \etal 2000b).
The planet has a mass $M_p=0.62 M_{\rm Jup}$ and orbits the star with a
period of 3.52 days with semimajor axis $a=0.047$ AU. Subsequent
observations of the transit with the STIS
camera on board HST have provided highly precise photometry of the
lightcurve during the transit (Brown \etal 2001), with about 300 flux
measurements of an accuracy close to $10^{-4}$ magnitudes. The shape
of the lightcurve has been used to determine the radius of the planet,
$R_p=1.35 R_{Jup}$.

  As discussed by Brown \etal (2001), the determination of transit
lightcurves of such high precision opens up many opportunities for
detecting other phenomena that could induce small changes in the
lightcurve, such as the presence of satellites or rings around the
planet. As more transits by other planets are discovered and observed
with high precision, the opportunities to observe these effects will
rapidly increase. This paper considers the effect of perturbations of
the planet's orbit due to the quadrupole moment of the star or to
a second planet. These perturbations should cause a precession of
the periastron of an eccentric orbit, and a precession of the line
of nodes if the orbital plane is not coincident with the stellar
equator, or the plane of the second planet. The precession of the
periastron and the line of nodes should result in long-term variations
of the period and the duration of the transits. We will
compute these two effects and discuss their detectability with future
space missions.

\section{Precession of the orbital plane}

  We consider first the effect of the precession of the orbital plane in
the simple case where the orbit is circular. The orbital angular
momentum of the planet is $L_p=M_p\, n\, a^2$, where $n =
(GM_s/a^3)^{1/2}$ is the mean motion (or orbital angular frequency), and
$M_s$ is the mass of the star. We consider two possible causes for the
precession of the orbital plane: the quadrupole moment of the star,
which for solid body rotation has angular momentum $L_s \simeq k^2 M_s
R_s^2 \omega_s$ (where $R_s$ is the stellar radius, $\omega_s$ the
angular frequency of rotation, and $k^2\simeq 0.1$ for main-sequence
stars; e.g., Ford, Rasio, \& Sills 1999), or a second planet, with
angular momentum $L_2 = M_2\, n_2\, a_2^2$, where we will assume $a_2 >
a$. The precession occurs relative to the plane perpendicular to the
total angular momentum vector ${\bf L}_t={\bf L}_p + {\bf L}_{(s,2)}$,
which will be referred to as the mean plane hereafter. The subindex
$(s,2)$ means that we are considering either the stellar rotation or a
second planet.

  Let $i$ be the inclination of the orbital plane relative to either the
stellar equator or the plane of the second planet, and $i_p$ and
$i_{(s,2)}$ the inclinations of these two planes relative to the mean
plane. Obviously, $i=i_p+i_{(s,2)}$, and $L_p\sin i_p = L_{(s,2)} \sin
i_{(s,2)}$. We define the x-axis as the intersection of the mean plane
and the plane perpendicular to the line of sight, and $\beta$ as the
angle between the mean plane and the line of sight. The line of nodes is
the intersection of the orbital plane of the planet and the mean plane,
and forms an angle $\Omega$ relative to the x-axis. The orbital
precession consists of the rotation of the angle $\Omega$, with a
precession angular frequency $\dot \Omega$.

\subsection{Precession induced by the stellar quadrupole moment}

  The potential of a star expanded up to its quadrupole moment is
given by
\begin{equation}
\phi = - {GM_s \over r} + {J_2\over 2}\, {GM_s R_s^2\over r^3}\,
(3\sin^2\theta - 1) ~,
\label{pots}
\end{equation}
where $R_s$ is the equatorial radius of the star, $r$ is the distance
to the center, $\theta$ is the angle relative to the equatorial plane,
and $J_2$ is the quadrupole moment. Estimates of the quadrupole moment
of the Sun are quite uncertain and over the range $10^{-7}$ to
$10^{-6}$ (Godier \& Rozelot 1999; Rozelot, Godier, \& Lefebvre 2001).
For self-similar main-sequence stars,
$J_2 \propto \omega_s^2 R_s^3/M_s$. In the case of the star HD209458,
the spectroscopically measured rotational velocity indicates a
rotation slightly faster than the Sun (Queloz \etal 2000). We will
use a fiducial value $J_2= 10^{-6}$ in this paper.

  To evaluate the time-averaged torque acting on the planet's orbit
as a secular perturbation, we calculate the potential of a ring of
mass $M_p$ in the quadrupole potential of the star,
\begin{equation}
V = {GM_s M_p R_s^2 J_2\over 2 a^3}\,
\int_0^{2\pi} {d\varphi \over 2\pi} (3\sin^2\varphi \sin^2 i - 1) =
{GM_s M_p R_s^2 J_2\over 2 a^3}\, \left({3\over 2} \sin^2 i - 1 \right) ~.
\end{equation}
The torque is
\begin{equation}
\tau = - dV/di = - {3 GM_s M_p R_s^2 J_2\over 4 a^3} \sin 2 i ~.
\end{equation}
Since the component of the planet's angular momentum that is changing
is $L_p \sin i_p$, the precession frequency is
$\dot\Omega = \tau/ (L_p \sin i_p)$. We will see below that the
quantity determining the change in the transit width is
$\dot\Omega \sin i_p$, which is
\begin{equation}
\dot\Omega \sin i_p = {\tau \over L_p} = n\, {R_s^2\over a^2}
{3J_2\over 4} \sin 2i ~.
\label{precs}
\end{equation}

\subsection{Precession induced by a second planet}

  The presence of a second planet will in general cause perturbations on
all the orbital elements. When averaged over long timescales, the
perturbations are classified as secular or resonant, and can be
calculated with Lagrange's planetary equations (e.g., Murray \& Dermott
1999, \S 6). Here we will consider only the secular precession of the
line of nodes, in the most simple case when both the perturbing and
perturbed planets are on circular orbits, which is easily obtained by
replacing the planets by uniform rings of mass. Moreover, we will use
the approximation $a_2 \gg a$. The case $a_2 < a$ is less interesting,
because the most accurate measurements of transits will be for planets
very close to their stars; however, the precession effects caused by a
planet interior to the transiting one can be calculated similarly to the
exterior case.

  The potential of a ring of radius $a_2$ and mass $M_2$, as a function
of the cylindrical radius $r$ on the plane of the orbit and the height
$z$ above the plane, is, up to second order in $r/a_2$ and $z/a_2$,
\begin{equation}
\phi= {-GM_2\over a_2} \left( 1 + {r^2\over 4a_2^2} - {z^2\over 2 a_2^2}
\right) ~.
\label{potp}
\end{equation}
The potential energy of interaction of this ring with another ring due
to the inner planet, of mass $M_p$ and radius $a$, and an inclination
angle $i$ between the two orbital planes, is
\begin{equation}
V= {-GM_p M_2\over a_2} \left( 1 + {a^2\over 4a_2^2} -
{3a^2\over 8 a_2^2}\sin^2 i
\right) ~,
\end{equation}
and the torque is
\begin{equation}
\tau = -dV/di = - {3 GM_p M_2 a^2 \over 8 a_2^3} \sin 2i ~.
\end{equation}
Just as before, the precession frequency is
\begin{equation}
\dot\Omega \sin i_p = {\tau\over L_p} = n {M_2\over M_s}\,
{a^3 \over a_2^3}\, {3\over 8} \sin 2i ~. 
\label{precp}
\end{equation}

\subsection{Effect on the transit duration}

  The properties of the transit lightcurve are determined by the
angle $\alpha$ between the orbital plane of the planet and the line of
sight. Simple spherical trigonometry shows that
\begin{equation}
\sin\alpha = \sin i_p \cos\beta \cos\Omega - \cos i_p \sin\beta ~,
\end{equation}
where $\beta$ is the angle between the mean plane and the line of
sight. The duration of the transit is
\begin{equation}
t_d = {2(R_s+R_p)\over n\, a}\, \cos\gamma ~,
\end{equation}
where the angle $\gamma$ is related to the impact parameter $b$ of the
transit by $(R_s+R_p) \sin\gamma = b = a \sin\alpha$. The relevant
question to determine the observability of the precession of the orbital
plane is if the rate of change of $t_d$ can be measured over a
reasonable observing time baseline $\Delta t_{obs}\sim 10$ years.
We will see that the typical value of the precession period is much
longer than 10 years, so only the time derivative $dt_d/dt$ matters:
\begin{equation}
{dt_d\over dt} = t_d\, {a\over R_s+R_p}\, {\sin\gamma\over \cos^2\gamma }
 \, {d\alpha \over dt} = t_d \, {a\over R_s+R_p}\,
{\sin\gamma\over \cos^2\gamma }\, \dot\Omega\, \sin i_p\,
\cos\beta\, \sin\Omega ~.
\label{tdd0}
\end{equation}
Using $n\, t_d = 2 (R_s+R_p)\cos\gamma/a$, we can reexpress this
equation as
\begin{equation}
{dt_d\over dt} = {\dot\Omega \sin i_p \over n}\, 2 \tan\gamma\,
\cos\beta\, \sin\Omega ~.
\label{tdd}
\end{equation}

  We can now estimate the value of $\dot\Omega \sin i_p$ and see if the
variation of the transit duration would be observable. A quadrupole
moment of the star $J_2\sim 10^{-6}$, with $a\sim 10 R_s$, and $\sin 2i
\sim 0.1$, yields a precession frequency $\dot\Omega \sin i_p \sim
10^{-9} n$, from equation (\ref{precs}). In the case of the perturbation
by a planet, if its mass is $M_2\sim 10^{-3} M_s$ (i.e., a gas giant),
the effect of the planet would be comparable to that of the quadrupole
for an orbital size $a_2\simeq 30 a$. For a 51-Peg type perturbed planet
with $a\simeq 0.05$ AU, the second Jupiter-mass planet would be at
$a_2\simeq 1.5$ AU. Such a massive planet should also be detected from
the Doppler measurements in any case. A more interesting case is if
there is a lower mass planet (not detectable by Doppler measurements)
on a smaller orbit. For example, for an Earth-like planet, with
$M_2\sim 3\times 10^{-6} M_s$, at $a_2 \sim 5 a \simeq 0.25$ AU, and
$\sin 2i\sim 0.1$, the precession frequency is also $\dot\Omega \sin i_p
\sim 10^{-9} n $.
Therefore, from equation (\ref{tdd}) (assuming that the trigonometric
factors are of order unity), in these cases we should expect
a time variation of the transit duration $dt_d/dt \sim 10^{-9}$.

  What is the highest accuracy to which we can measure $dt_d/dt$ ?
In the STIS observations of Brown \etal (2001) of HD209458, the flux
varies by $\sim$ 1\% during a time $t_c \simeq 0.01$ days at the
beginning and the end of the transit. Each one of their flux
measurements is obtained from a 60 s integration and has a relative
accuracy of $\sim 10^{-4}$. Therefore, each data point taken during
the falling or rising part of the lightcurve gives us the starting or
ending time of the transit to an accuracy $\sim (10^{-4}/0.01) t_c$,
or $\sim 10$ seconds. Brown \etal have measured nearly 100 data
points in the rapidly varying part of the lightcurve, so their present
data should allow to determine the transit duration to an accuracy
approaching 1 second. Repeating the measurement 3 years later would
yield $dt_d/dt$ to an accuracy of $10^{-8}$.

  The star HD209458 has magnitude $V=7.64$, implying that the number of
photons received by HST over a 60 s integration time is $\sim 10^9$.
Therefore, the photometric accuracy achieved by Brown \etal could be
improved by only a factor $\sim 3$ by reaching the photon shot noise
limit; alternatively, a telescope aperture of $0.7$ m could reach the
same photometric accuracy with improvements in the detection efficiency.
A dedicated mission observing all the $\sim 1000$ transits over a
period of 10 years of stars similar to HD209458 with the same accuracy
as Brown \etal would yield $dt_d/dt$ to an accuracy $\sim 10^{-9.5}$.

  We can easily change the parameters of the examples discussed
previously to consider cases where the orbital perturbations would be
detectable with the more easily achievable accuracy of $dt_d/dt\simeq
10^{-8}$. For example, a perturbing planet with $M_p=3\times 10^{-5}
M_s$, $a_2/a=4$, and $\sin 2i=0.3$, would be detectable at the
5-$\sigma$ level.

\section{Precession of the periastron}

  The orbits of the 51-Peg type planets tend to be circular, as
expected owing to tidal dissipation in the planet, which
circularizes the orbit on a timescale (Goldreich \& Soter 1966)
\begin{equation}
\tau_e = {4\over 63}\, Q\, n^{-1}\,{M_p\over M_s}
\left( {a\over R_p} \right)^5 ~,
\end{equation}
where the factor $Q$ is inversely proportional to the dissipation
rate. For the planet Jupiter, $Q\sim 10^5$ has been found
observationally from the tidal effect on Jupiter's satellites
(Ioannou \& Lindzen 1993). Assuming the same $Q$ for HD209458,
for which case $M_p/M_s\simeq 5\times 10^{-4}$ and $a/R_p\simeq 75$,
we find $\tau_e\sim 10^7$ years. How the parameter $Q$ may vary among
different planets is highly uncertain, because the origin of the
dissipation is not well understood (e.g., Murray \& Dermott 1999,
\S 4.13). So, while internal dissipation in the planets is a
plausible explanation for the orbital circularity of the closest
planets, it is possible that some of the closest planets may have
eccentric orbits. In addition, the extreme sensitivity of the time
$\tau_e$ to $a/R_p$ implies that planets that are only slighly further
out from their star can have significant eccentricities even if $Q$ is
constant. Orbital eccentricities can also be excited by planetary
companions (e.g., Rivera \& Lissauer 2000). In fact, among known
extrasolar planets, high eccentricities are common when $a > 0.1$ AU
(Butler \etal 2000), and the planet around HD217107 has $a=0.072$ AU and
$e=0.14\pm 0.05$ (Fischer \etal 1999). We can
expect that transits of planets in eccentric orbits will be discovered
in the near future.

  We note here that tidal dissipation in the star can also circularize
the orbit, and make the orbit coplanar with the stellar equator.
However, the timescale for dissipation in the star is much longer than
the age of the system, and is in any case similar to the timescale for
orbital decay (Rasio \etal 1996; Zahn 1977).

\subsection{Precession rates}

  The periastron of an eccentric orbit should precess due to three
effects: the relativistic precession, and the same two effects
considered previously, the stellar quadrupole moment and the
perturbations from other planets.

  The relativistic precession rate is given by (e.g., Landau \&
Lifshitz 1951)
\begin{equation}
\dot\varpi = n\, {3\over 1-e^2}\, \left( {n\, a\over c} \right)^2 ~,
\end{equation}
where $e$ is the eccentricity, and we denote the precession of the
periastron as $\dot\varpi$.
For an orbital period of ten days, $a\sim 0.1$ AU, and small
eccentricity, this gives
$\dot\varpi\simeq 4\times 10^{-7} n \simeq 10^{-4} \yr^{-1}$.

  For the stellar quadrupole moment, we assume here that the orbital
plane coincides with the stellar equator. The potential of the star
is given by equation (\ref{pots}) with $\theta=0$. The precession rate for
small eccentricities is easily obtained using the epicycle
approximation. The orbital angular frequency is
$n^2 =(1/r) d\phi/dr$, and the epicycle frequency is
(e.g., Binney \& Tremaine 1987)
\begin{equation}
\kappa^2= r {d n^2\over dr} + 4 n^2 = n^2 -
{3GM_s R_s^2 J_2 \over r^5 } ~.
\end{equation}
The precession frequency is
\begin{equation}
\dot\varpi = n -\kappa = {3 J_2 R_s^2 \over 2 a^2}\, n  ~,
\label{pres}
\end{equation}
where we have substituted $a=r$ and have used $\dot\varpi \ll n$.
Typically, $J_2\simeq 10^{-6}$ and $R_s \lesssim 0.1 a$, so
$\dot\varpi \lesssim 10^{-8} n$. Except in rapidly rotating stars, the
effect of the stellar quadrupole is always much less than the
relativistic precession.

  For the effect from a second planet, repetition of the same method
with the potential in equation (\ref{potp}) at z=0 yields a precession
frequency
\begin{equation}
\label{prep}
\dot\varpi = {3 M_2 a^3 \over 4 M_s a_2^3}\, n ~.
\end{equation}
For an Earth-like planet with $M_2/M_s\simeq 3\times 10^{-6}$, and
$a_2\simeq 2a$, the precession rate is $\dot\varpi\simeq 3\times 10^{-7}
n$, comparable to that from the relativistic precession.
All the three effects we have discussed cause an advance of the
periastron, and therefore the total precession is the sum of the
three effects.

\subsection{Detectability of the periastron precession}

  The periastron precession causes both the period and the duration
of the transits to change. We discuss first the variation of the
period.

  In the epicycle approximation (valid for small eccentricities), the
eccentric orbit of the planet can be described by the motion along an
epicycle relative to a circular orbit, with coordinates $x(t)$ along
the inward radial direction and $y(t)$ in the backward tangential
direction along the orbit. For an orbit in a
quasi-Keplerian potential (where the precession of the periastron
is small), we have (Binney \& Tremaine 1987):
\begin{equation}
x(t)= ae \cos(\kappa t + \psi_0) ~;
\end{equation}
\begin{equation}
y(t)= - 2ae \sin(\kappa t + \psi_0) ~.
\label{epiy}
\end{equation}
The central time of the transits is determined by $y(t)$ at the time of
the transit. Between two successive transits, the epicycle phase $\psi =
\kappa t + \psi_0$ changes by $\Delta \psi = 2\pi (\dot\varpi/n)$, and
therefore $y(t)$ changes by $\Delta y = 2ae\, 2\pi (\dot\varpi/n)\,
\cos(\kappa t + \psi_0)$. Thus, the observed transit period, $P_t$, is
(to first order in $e$)
\begin{equation}
P_t= {2\pi\over n} \, \left(1-{\Delta y \over 2\pi a} \right) =
{2\pi\over n} \, \left(1-2e{\dot\varpi\over n}\, \cos\psi
\right) ~.
\end{equation}

  A first possible method to detect the periastron precession is to
compare the transit period $P_t$ with the true orbital period
$P=2\pi/n$ determined from the Doppler measurements.
The accuracy of the Doppler measurements are at present
$\epsilon_D \sim 0.1$ times the velocity variation amplitude, and they
are unlikely to improve very much due to limitations set by photospheric
turbulence. Hence, the orbital period can be determined to an accuracy
$\Delta P/P\simeq \epsilon_D P/(2\pi N^{1/2} t_0)$, where $N$ is the
number of Doppler measurements and $t_0$ is the total time of
observation. For $\epsilon_D=0.1$, $N=1000$ and $P/t_0=10^{-3}$, the
accuracy achieved is $\Delta P/P \sim 10^{-6}$, allowing detection
only for a planet with $M_2/M_s\gtrsim 10^{-4}$ if $e\sim 0.2$ and
$a_2/a\simeq 2$.

  A second possibility is to detect the change of $P_t$ with time:
\begin{equation}
{dP_t\over dt} = 4\pi\, e\, \left(\dot\varpi\over n\right)^2\,
\sin\psi ~.
\label{ptd}
\end{equation}
If the time of each transit can be measured to an accuracy of $1$
second, the accuracy of the period after observing $N=1000$ transits
would be $1{\rm s}/N^{3/2}\sim 10^{-4.5}$ s, and for an observing time
of 10 years the accuracy of $dP_t/dt$ would reach $\sim 10^{-13}$. For
$e\sim 0.1$, this allows for a 10-$\sigma$ detection of a planet with
$M_2/M_s=10^{-5}$ and $a_2=2a$. 

  Next we evaluate the change in the duration of the transit, which is
caused by the variation of the planet's velocity during transit as the
periastron precesses. From equation (\ref{epiy}), the velocity of the
planet at transit is $v= n a (1-2e\cos\psi)$ (notice that for the
purpose of computing this velocity we can use the approximation
$\kappa\simeq n$), so the duration is
\begin{equation}
t_d = {2 (R_s+R_p)\cos\gamma \over v} \simeq 2 \,
{(R_s+R_p)\over na}\, \cos\gamma\, (1+2e \cos\psi) ~,
\end{equation}
and its time derivative is
\begin{equation}
{dt_d\over dt} = 4e\, {\dot\varpi\over n}
{(R_s+R_p)\over a}\, \cos\gamma\, \sin\psi ~.
\label{tdde}
\end{equation}
As mentioned previously, the perturbation by an Earth-like planet with
$a_2=2a$ implies a periastron precession rate $\dot\varpi/n \sim
3\times 10^{-7}$; for $R_s/a\sim 0.05$, $4e\sim 1$, we infer
$dt_d/dt \sim 10^{-8}$, which is not difficult to detect as discussed
in \S 2.3. From equations (\ref{precp}) and (\ref{tdd}),
we see that the transit duration change due to the precession of the
orbital plane would be comparable if $\sin 2i\simeq 0.1$.

  In principle, the change in transit duration due to the precession
of the periastron and the lines of nodes can be distinguished by the
effect on the shape of the lightcurve. Precession of the periastron
changes only the velocity, so the shape of the lightcurve does not
change. However, precession of the line of nodes changes the impact
parameter of the transit. Of course, separating the two effects requires
a higher photometric accuracy, and will increase the measurement errors
unless the effect of the periastron precession is also measured from the
change in the transit period. The most favorable case would be when the
planet transits over the edge of the star, when the lightcurve shape has
the fastest change with a variation of the impact parameter.

  It is simple to see which one of the two measurements, the change
in the period or the duration of the transit, can yield the highest
accuracy to measure the ratio $\dot\varpi/n$. The period derivative
is always measured to an accuracy better than the transit duration
derivative by a factor $P/t_0$, where $t_0$ is the time of
observation. Equating this to the ratio of equations (\ref{ptd})
and (\ref{tdde}), we find that the transit duration change yields
a higher accuracy until the time of observation is
$t_0 < \dot\varpi^{-1}\, 2(R_s+R_p)\cos\gamma/a$,
and the period change gives the most accurate determination after that.
For a close-in planet with $2R_s/a\simeq 0.1$, and an Earth-mass
perturbing planet with $a_2/a\simeq 2$, we find
$\dot\varpi^{-1}\sim 3\times 10^4$ years. Therefore, the transit
duration change should yield the periastron precession rate with the
highest accuracy, except for substantially more massive perturbing
planets.

\section{Conclusions}

  Accurate photometry of transit lightcurves of extrasolar planets
allows the detection of slow orbital perturbations which affect the
lightcurve. Here, we have considered the perturbation due to a stellar
quadrupole moment or to a second planet as possible causes for the
precession of the line of nodes and of the periastron of the orbit.
Both types of precession cause a time variation of the duration of
the transit, and the periastron precession also causes a variation of
the period. The rate of this variation is given by equations
(\ref{tdd}), (\ref{tdde}), and (\ref{ptd}).

  The quadrupole moment of the star is important for the close-in
planets, at $a\sim 10 R_s$. These planets are likely to be on
circular orbits and therefore the precession of the line of nodes
should usually be the most important effect. From observations of
the stellar spectrum one can infer the projected rotational velocity
of the star, and monitoring the CaII flux can also reveal the rotation
period (Henry \etal 2000a). In addition, spectroscopic variations of the
star during the transits can in principle be used to measure the angle
$\Omega$ of the line of nodes (Quelloz \etal 2000). If all these
quantities are measured, it should be possible to infer from the
precession rate of the orbital plane, the quadrupole moment of the
star $J_2$, which is sensitive to the state of rotation of the inner
parts of the star.

  The precession induced by other planets will be added to that
caused by the stellar quadrupole, and only the sum of both effects
is observable. However, once the stellar rotation is measured the
effect of the quadrupole is predictable to some extent, and it should
be negligible compared to the effect of another planet in many cases
(for large $a$ or $M_2$).

  The precession induced by a second planet depends only on the
quantities $(M_2/M_s)(a/a_2)^3$, and $\sin 2i$ (see eqs.\ [\ref{precp}]
and [\ref{prep}]). In principle, by measuring both the orbital and
periastron precession from the change in the duration and the shape
of the lightcurve, or from measuring the change in the period as well,
both of these quantities could be determined assuming that there is only
one perturbing planet. Even then, the mass and semimajor axis of the
perturbing planet cannot be separately inferred, and the detected
precession could be the result of a low-mass planet close to the
transiting planet, or a more massive planet further away, or a disk of
planetessimals.

  In the examples mentioned previously in this paper, we have
considered perturbations by planets of low enough mass that they would
not be detected via the Doppler measurements of the star. However,
transits could also allow the detection of perturbations by other known
planets. Several cases of systems with two or three planets have now
been discovered, some of which seem to be on resonant orbits (Marcy
\etal 2001). Measuring the orbital and periastron precessions in one of
these systems, if they were to show transits, could allow the
determination of their relative orbital inclination. For a case with
$M_2/M_s\sim 10^{-3}$, $a_2/a=2^{2/3}$, the precession periods may be
as short as hundreds of years, making it possible that a whole series
of transits may be observed over a reasonable time of $\sim 10$ years
as the line of nodes precesses, and enabling very accurate measurements
of the orbital perturbations. A more general analysis than that
presented here would be required to compute the precession rates when
the two planets are not coplanar, are both on eccentric orbits, and
are possibly resonant, including also the effect of changes of the
eccentricity on the transit period and duration.

\acknowledgements

  I am grateful to David Weinberg and the Ohio State astronomy graduate
students Khairul Alam, Chris Burke, Julio Chanam\'e, Frank Delahaye,
Dale Fields, Susan Kassin, Jennifer Marshall, Chris Onken,
James Pizagno, Jeremy Tinker, and Zheng Zheng, for a wonderful class
on the precession of the orbital plane induced by a planet, which made
me think of the interesting case of a perturbing Earth-mass planet.
I thank the Institute for Advanced Study for their hospitality when
this work was being completed.


\newpage
\vskip -0.2cm
\begin{thebibliography}{}

\bibitem[]{} Binney, J., \& Tremaine, S. 1987, {\it Galactic Dynamics}
(Princeton University Press).
\bibitem[]{} Brown, T. M., Charbonneau, D., Gilliland, R. L., Noyes, R. W.,
\& Burrows, A. 2001, ApJ, 552, 699
\bibitem[]{} Butler, R. P., Marcy, G. W., Vogt, S. S., \& Fischer, D. A.
2000, in {\it Planetary Systems in the Universe}, IAU Symp. 202
\bibitem[]{} Charbonneau, D., Brown, T. M., Latham, D. W., \& Mayor, M.
2000, ApJ, 529, L45
\bibitem[]{} Fischer, D. A., Marcy, G. W., Butler, R. P., Vogt, S. S.,
\& Apps, K. 1999, PASP, 111, 50
\bibitem[]{} Ford, E., Rasio, F. A., \& Sills, A. 1999, ApJ, 514, 411
\bibitem[]{} Godier, S., \& Rozelot, J.-P. 1999, A\& A, 350, 310
\bibitem[]{} Goldreich, P., \& Soter, S. 1966, Icarus, 5, 375
\bibitem[]{} Henry, G. W., Baliunas, S. L., Donahue, R. A., Fekel, F. C.,
\& Soon, W. 2000, ApJ, 531, 415
\bibitem[]{} Henry, G. W., Marcy, G. W., Butler, R. P., \& Vogt, S. S.
2000b, ApJ, 529, L41
\bibitem[]{} Ioannou, P. J., \& Lindzen, R. S. 1993, ApJ, 406, 266
\bibitem[[]{} Landau, L. D., \& Lifshitz, E. M. 1951,
{\it The Classical Theory of Fields} (Cambridge: Addison-Wesley).
\bibitem[]{} Marcy, G. W., Butler, R. P., Fischer, D., Vogt, S. S.,
Lissauer, J. J., \& Rivera, E. J. 2001, ApJ, 556, 296
\bibitem[]{} Murray, C. D., \& Dermott, S. F. 1999, {\it Solar System
Dynamics} (Cambridge University Press)
\bibitem[]{} Queloz, D., Eggenberger, A., Mayor, M., Perrier, C.,
Beuzit, J. L., Naef, D., Sivan, J. P., \& Udry, S. 2000, A\& A, 359, L13
\bibitem[]{} Rasio, F. A., Tout, C. A., Lubow, S. H., Livio, M. 1996, ApJ,
470, 1187
\bibitem[]{} Rivera, E. J., \& Lissauer, J. J. 2000, ApJ, 530, 454
\bibitem[]{} Rozelot, J.-P., Godier, S., \& Lefebvre, S. 2001, Sol. Phys.,
198, 223
\bibitem[]{} Zahn, J.-P. 1977, A\& A, 57, 383; erratum 67, 162

\end{thebibliography}
\end{document}